\documentclass[aps,jcp,twocolumn]{revtex4-1}
\usepackage{amsmath}
\usepackage{amsfonts}
\usepackage{graphicx}
\usepackage{hyperref}

\newcommand*{\bmuex}[1]{\beta\mu^\text{ex}_{#1}}
\newcommand*{\bmuid}[1]{\beta\mu^\text{id}_{#1}}
\newcommand*{\inner}[1]{\ensuremath{ \left\langle #1 \right\rangle }}
\newcommand*{\pd}[2]{\ensuremath{\frac{\partial{#1}}{\partial{#2}}}}

\newcommand*{\avg}[1]{\ensuremath{\left\langle{#1}\right\rangle}}

\newcommand*{\erfc}{\operatorname{erfc}}

\newcommand*{\Tr}[1]{\operatorname{Tr}\left[#1 \right] }
\begin{document}

\title{ Extension of Kirkwood-Buff Theory to the Canonical Ensemble}
\author{ David M. Rogers}
\affiliation{ University of South Florida, 4202 E. Fowler Ave., CHE 205, Tampa,
FL 33620, USA}
\email{ davidrogers@usf.edu}

\date{\today}

\begin{abstract}
   Kirkwood-Buff (KB) integrals are notoriously difficult to converge from a canonical
simulation because they require estimating the grand-canonical radial distribution.
The same essential difficulty is encountered when attempting to estimate the direct
correlation function of Ornstein-Zernike theory by inverting the pair correlation functions.
We present a new theory that applies to the entire, finite, simulation volume,
so that no cutoff issues arise at all.
The theory gives the direct correlation function for closed systems, while
smoothness of the direct correlation function in reciprocal
space allows calculating canonical KB integrals via a well-posed extrapolation to the origin.
The present analysis method represents an improvement over previous work because
it makes use of the entire simulation volume and its convergence can be accelerated
using known properties of the direct correlation function.
Using known interaction energy functions can make this extrapolation
near perfect accuracy in the low-density case.
Because finite size effects are stronger in the canonical than the grand-canonical
ensemble, we state ensemble correction formulas for the chemical potential
and the KB coefficients.
The new theory is illustrated with both analytical and simulation results on the
1D Ising model and a supercritical Lennard-Jones fluid.  For the latter, the
finite-size corrections are shown to be small.
\end{abstract}


\maketitle

\section{ Introduction}

  Kirkwood and Buff's theory of solutions exploits a Gibbs relation to
determine derivatives of activities from fluctuations in particle number at constant volume,
\begin{equation}
\left( \pd{\beta \mu_\alpha}{n_\gamma}\right)_{n_\eta,V,T}
  = \left[\avg{n_{\alpha'} n_{\gamma'} | \mu,V,T} - V^2 \rho_{\alpha'} \rho_{\gamma'} \right]_{\alpha\gamma}^{-1},
\end{equation}
with the notation $\rho_\alpha = \avg{n_\alpha | \mu, V, T}/V$, the average number of molecules
of type $\alpha$ in the grand-canonical ensemble (where $\mu, V, T$ are given),
and using $\mu \in \mathbb R^\nu$ to
denote a vector of chemical potential values and $n \in \mathbb N^\nu$, a
vector of molecule numbers for all $\nu$ solution components.
The inverse indicated above is the Moore-Penrose pseudoinverse of the $\nu\times\nu$ matrix of
number fluctuations.
KB theory is exact even when applied to periodic simulation volumes with finite unit cells,
but requires an {\em open} system (i.e. data from the $\mu,V,T$ ensemble).
In this work, we consider only such finite unit cells and use the additional
assumption that no fixed external potential is present that breaks the
translational symmetry.  We introduce a theory that applies equally to open and closed systems
and then investigate its zero-frequency limit (which is only rigorous for the open system).
Also, we maintain constant temperature throughout
and hence omit it in further notation.

  KB theory is commonly cast in terms of the grand-canonical radial distribution function,
\begin{equation}
g_{\alpha\gamma}(r | \mu, L) \equiv \frac{
\avg{\hat \rho_\alpha(r) \hat \rho_\gamma(0) | \mu, L} - \delta_{\alpha\gamma} \delta(r) \rho_\alpha
  }{
\rho_\alpha \rho_\gamma
},\label{e:g}
\end{equation}
whose integral reports on number fluctuations in the grand-canonical ensemble.
Here, $r \in \mathbb R^3 / L$ is a coordinate vector in the periodic unit cell
with lattice vectors given as rows of a 3$\times$3 matrix, $L$,
and
\begin{equation}
\hat \rho_\alpha(r) \equiv \sum_{i=1}^{n_\alpha} \delta(r - r_{\alpha,i})
\end{equation}
 is a sum of delta-functions located at the centers
of mass for all particles of type $\alpha$ (so that
$\int_V \hat \rho_\alpha(r) dr = n_\alpha$ and $\rho_\alpha = \avg{\hat \rho_\alpha(0) | \mu, L}$).
Directly integrating Eq.~\ref{e:g} gives Kirkwood and Buff's Eq.~7,\cite{kb}
confirming that it applies to {\em finite} volumes with the entire volume as the domain of integration.

  A recurring problem is that KB integrals estimated from closed systems (i.e. those under
NVT or NPT conditions) contain substantial finite-size effects.\cite{nmatu96,pgang13}
Proposed corrections fall into two categories: those aimed at computing number
density fluctuations directly in small sub-volumes,\cite{mrove90,sschn11,rcort16}
and those addressing the cutoff error in truncating the KB integral
at a finite maximum radius.\cite{nmatu96,pkrug13,pgang13}
Both categories recognize that {\em two} independent corrections are necessary.
First, the sub-volume or truncation radius is typically extrapolated
to infinite size.  Second, the radial distribution (or equivalently the number correlations)
themselves must be corrected for the effect of using a closed system.
These corrections were defined as implicit and explicit by
Salacuse et. al.~\cite{jsala96,jsala96b}, who derived and compared
calculations of the closed system correction.
However, the finite truncation radius remained as a source of ambiguity,
which more recent works have addressed with renewed interest.\cite{pkrug13}

  In this work, we re-introduce the structure factor as the preferred method for computing both the
direct correlation function and the KB integral as its zero-frequency limit.
This corrects a minor defect in Ref.~\citenum{jsala96} by using the unique vectors in reciprocal space
that correspond to a finite simulation volume, and yields exact direct correlation functions
corresponding to the canonical ensemble simulated.
For the problem of estimating the thermodynamic limit, our method improves on earlier works as it
remains invariant to shifts of $g(r|n, L)$ by a constant,
and makes use of the entire available simulation volume.
In the process, we find two major conclusions.
First, extrapolation of the direct correlation function for closed systems
to zero frequency gives good estimates
of the second derivatives of the Helmholtz free energy.
Second, both closed (canonical) and open (grand-canonical) KB integrals
can be estimated under near-critical conditions from box lengths on the order
of 5 to 10 times the correlation length.

\section{ Theory}

\subsection{ Definitions}
  We introduce the structure factors using Fourier transforms of the density operators,
\begin{align}
\hat S_\alpha(m) &\equiv \mathcal F[\hat \rho_\alpha](m) \equiv \int_V \hat \rho_\alpha(r) e^{-2\pi i m\cdot r} \; dr
.
\intertext{Explicit formulas for these transforms are,}
\hat S_\alpha(m) &= \sum_{j=1}^{n_\alpha} e^{-2\pi i m\cdot r_{\alpha,j}}, \,
\hat \rho_\alpha(r) = \frac{1}{V} \sum_m \hat S_\alpha(m) e^{2\pi i m\cdot r}
,
\end{align}
where the sum over $m$ runs over the infinite reciprocal
lattice with vectors $m = L^{-1} u$ ($u\in\mathbb Z^3$),
and the volume of the periodic unit cell is $V = |L|$.
These sums can be computed efficiently using the fast Fourier
transform in molecular simulations.\cite{uessm95}
The results in this work were computed with our own
implementation of the method,
which is available with documentation in Ref.~\citenum{droge17d}.

  For any fixed-volume ensemble with partition function $Z(\Gamma,L)$
(e.g. $\Gamma = n,T$ for canonical or $\Gamma = \mu,T$ for grand-canonical),
and probability distribution $P(\tau | \Gamma,L)$ (where $\tau$
represents the combined positions and momenta of all molecules present),
define an extended ensemble that characterizes its response
to a set of externally applied fields, $\phi = \{\phi_\alpha\}$, using the partition function,
\begin{align}
\Theta(\phi, \Gamma, L) &= Z(\Gamma, L) \int  
     e^{\inner{\hat \rho, \phi}} \; P(\tau | \Gamma, L) d\tau \\
\inner{X,Y} &\equiv \sum_\alpha \int_V X_\alpha(r)^\dagger Y_\alpha(r) dr \notag \\
&= \frac{1}{V} \sum_\alpha \sum_m \mathcal F[X_\alpha](m)^\dagger \mathcal F[Y_\alpha](m)
\label{e:inner}.
\end{align}
The last two lines define a convenient notation for inner products
between two functions of $r$ in the unit cell and state the appropriate
Poisson summation formula.\cite{lax}
We also define the function,
\begin{align}
\tilde \phi_\alpha(m) &\equiv \mathcal F[\phi_\alpha](m) / V \\
\intertext{to absorb the normalization constant in Eq.~\ref{e:inner}.  This way,}
\inner{\hat \rho, \phi} &= \sum_\alpha \sum_m \hat S_\alpha^\dagger(m) \tilde \phi_\alpha(m), \\
\text{and } \tilde \phi_\alpha(0) / \beta &= \mu_\alpha,
\end{align}
the chemical potential of species $\alpha$.
The first and second derivatives of $\ln \Theta$ generate the densities and correlations,
\begin{align}
\pd{\ln \Theta}{\tilde \phi_\alpha(m)^\dagger} &= \avg{ \hat S_\alpha(m)} \equiv S_\alpha(m)
\label{e:S} \\
\pd{^2 \ln \Theta}{\tilde \phi_\alpha(m)^\dagger \tilde \phi_\gamma(m')}
   &= \pd{S_\alpha(m)}{\tilde \phi_\gamma(m')}
   = V \delta_{m,m'} Q_{\alpha\gamma}(m) \label{e:corr} \\
Q_{\alpha\gamma}(m) &\equiv \frac{1}{V} \avg{ \Delta \hat S_\alpha(m) \Delta \hat S_\gamma(m)^\dagger } 
 \\
 &= \mathcal F[\avg{\hat\rho_\alpha(r) \hat\rho_\gamma(0)}](m) - \delta_{m,0}V\rho_\alpha\rho_\gamma, \label{e:Qg}
\end{align}
where $\Delta \hat S_\alpha(m) \equiv \hat S_\alpha(m) - S_\alpha(m)$.
All quantities in Eqns.~\ref{e:S}-\ref{e:Qg} are conditional on the macrostate $(\phi, \Gamma, L)$
and evaluated at points where $\phi$ is spatially uniform.
The vanishing of $m \ne m'$ cross-derivatives in Eq.~\ref{e:corr} can be proven
from translational symmetry in this case.

  The matrix $Q/\rho$ is the structure factor of scattering theory.\cite{jsala96}
It is connected to the Fourier transform of the radial distribution function
via Eqns.~\ref{e:Qg} and~\ref{e:g}.
It is also related to the integral theory of solutions via the definition,\cite{pdt6}
\begin{equation}
Q^{-1}_{\alpha\gamma}(r) = \delta(r) \pd{\beta \mu^\text{id}_\alpha}{\rho_\gamma} - c_{\alpha\gamma}(r)
,\label{e:OZ}
\end{equation}
of the Ornstein-Zernike direct correlation function, $c_{\alpha\gamma}(r)$.
Here, $\mu_\alpha^\text{id}$ is the chemical potential of an ideal gas,
whose derivative is $\partial \beta\mu_\alpha^\text{id}/\partial\rho_\alpha = 1/\rho_\alpha$.

\subsection{ Canonical Kirkwood-Buff Coefficients}

  It is well-known that the matrix inverse of Eq.~\ref{e:corr}
gives the shift in the potential required to maintain a small
change in the density profile appropriate for an ensemble with fixed $(S, \Gamma, L)$.
Eqns.~\ref{e:S}-\ref{e:Qg} apply exactly to both canonical ($\Gamma = n,T$)
and grand-canonical ($\Gamma = \mu,T$) ensembles.

  From this connection, the most relevant quantities to canonical KB theory
are the fit coefficients, $c_0$ and $c_2$, for
\begin{equation}
\lim_{m\to 0} Q_{\alpha\gamma}^{-1}(m | n,L) = \pd{\beta \mu^\text{id}_\alpha}{\rho_\gamma}
+ c_{0,\alpha\gamma}(0) + c_{2,\alpha\gamma} (2\pi m)^2 + O(m^4)
.\label{e:fit}
\end{equation}
There is no linear term since $Q$ must be an even function of $m$ by symmetry.
For a sufficiently large number of molecules of type $\alpha$, the first
coefficient provides the density dependence of their excess chemical potential,
\begin{equation}
\lim_{L /L_c \to\infty} c_{0,\alpha\gamma}(n, V) = \beta V \left(
\mu^\text{ex}_\alpha(n,V)
- \mu^\text{ex}_\alpha(n - \delta_\gamma,V)
\right)
,\label{e:dmuex}
\end{equation}
where we define
\begin{equation}
\beta\mu^\text{ex}_\alpha(n,V) \equiv -\ln \avg{ e^{-\beta \Delta U_\alpha} | n, V}
,
\end{equation}
to be the test particle insertion free energy in the canonical ensemble.\cite{pdt2}
The coefficient $c_{0,\alpha\gamma}$
also converges to $\partial \beta\mu_\alpha^\text{ex} / \partial\rho_\gamma$
in the thermodynamic limit.
Convergence to the particle insertion free energy
is faster though, because the Helmholtz free energy
can be rigorously stated as the sum of $n$ successive test particle insertions,
\begin{equation}
\beta A(n,V,T) = \beta A^\text{id}(n,V,T) + \sum_{j=0}^{n-1} \beta\mu^\text{ex}(j,V,T)
,
\end{equation}
while the ideal term is indistinguishable from $n \ln \rho$ after $n \ge 100$.

  As further justification of the limit, we invoke the well-known
result that density fluctuations must become uncorrelated as the
wavelength, $1/m$ grows much larger than the correlation length.\cite{jlebo67}
%, for real fluids with short-range interactions and away from phase transitions.
Thus, under these conditions $Q^{-1}(m)$ is continuous as $m$ approaches zero
and its extrapolation becomes easier as the
cell size increases and the range of interactions becomes shorter.

  Away from $m=0$, the matrix $Q$ reports on structural features that cannot be calculated
from bulk thermodynamic data.  The second coefficient, $c_2$, in Eq.~\ref{e:fit}
provides the correlation length for the $\alpha\gamma$ radial distribution,\cite{davis9}
\begin{equation}
L_c^{(\alpha\gamma)} = \sqrt{ \left |
Q_{\alpha\gamma}(0) c_{2,\alpha\gamma} \right |} \label{e:clen}
.
\end{equation}
The coefficient $c_2$ is generally negative for oscillatory decay
and positive for monotonic decay of the radial distribution, $g(r)$.

  The identification of Eq.~\ref{e:dmuex} as the proper limiting form
for the canonical ensemble allows us to carry through the rest of the
Kirkwood-Buff analysis unchanged.
All the other derived quantities of Kirkwood-Buff theory can be stated by noting that
the matrix $Q_{\alpha\gamma}(0|\mu, L)$ is identical to
$B_{\alpha\gamma}$ defined in Eq.~9 of Ref.~\citenum{kb}.
The supplementary information provides explicit expressions.
In the thermodynamic limit, the density-derivatives of the chemical potential
are the same for the canonical and grand-canonical ensembles.
However, at finite volume, the grand-canonical ensemble provides
observables closer to the thermodynamic limit than the canonical.

  From this identification, we conclude that the most effective method
for estimating KB integrals from canonical simulations is to use density
fluctuations over the entire simulation cell to estimate $Q(m)$ and then
to extrapolate the value of $Q(0)$.  Our numerical results below
indeed show this works, but caution that there are two sources
of error.  The first is due to the difficulty of extrapolating to $m=0$.
The second is due to finite size corrections, which become increasingly
important for successive density derivatives.
This work shows that the first type of error can be entirely eliminated
within the canonical ensemble when the simulation box size is larger than
about ten correlation lengths (which can be easily estimated using Eq.~\ref{e:clen}).

%The estimation of the thermodynamic limit suggested in Ref.~\citenum{pkrug13},
%is found to be satisfactory.
%
%It amounts to assuming $\pd{^2}{\rho^2} Q(m)$ in Eq.~\ref{e:finite}
%is relatively scale-independent.
%The estimation error sometimes {\em increases} when extrapolating to $m=0$.

\subsection{ Finite Size Corrections}

  Finite size corrections (which extrapolate a result at finite volume to
the thermodynamic limit) are more severe for the canonical
than the grand-canonical ensemble.  This section derives
relations connecting averages in the canonical and grand-canonical
ensembles.  It is usually assumed that the grand-canonical
value is close enough to the thermodynamic limit that this ensemble
correction is a good estimate of the finite size correction.

  The correction to the distribution function was given by Lebowitz~\cite{jlebo67},
\begin{equation}
\begin{split}
Q_{\alpha\gamma}&(m | \mu, L) - Q_{\alpha\gamma}(m | n, L) \\
  &= \frac{1}{2V} \sum_{\eta\zeta} Q_{\eta\zeta}(0 | \mu, L)
      \pd{^2}{\rho_{\eta} \partial \rho_{\zeta}} Q_{\alpha\gamma}(m | \mu, L)
  ,
 \end{split}
      \label{e:finite}
\end{equation}
which is correct to relative order $1/V$ unless $m=0$.
At the point $m=0$, $Q_{\alpha\gamma}(0 | n, L) = 0$, and the entire
contribution to $Q_{\alpha\gamma}(0 | \mu, L)$ comes from the second term.
Eq.~\ref{e:finite} was shown to match the thermodynamic
limit in Ref.~\citenum{jsala96b}.

  Using the same method, we derive in the appendix the correction
to the excess chemical potential valid to first order in $1/V$,
\begin{equation}
%\frac{ \avg{e^{-\beta\Delta U_\alpha} | \mu, L} }{ \avg{ e^{-\beta\Delta U_\alpha} | n, L} }
%= 1 -
\begin{split}
\beta \mu_\alpha^\text{ex}&(\mu,V) - \beta \mu_\alpha^\text{ex}(n,V) \\ &=
   \frac{1}{2V}\left(
         \frac{2}{\rho_\alpha} - Q^{-1}_{\alpha\alpha}(0)
         + \sum_{\gamma} \pd{Q^{-1}_{\alpha\gamma}}{\beta\mu_\gamma}(0)
    \right).
\end{split}
        \label{e:dmu}
\end{equation}
For a 1-component system, $Q(0) = \rho^2\kappa / \beta$ is proportional to the isothermal
compressibility, $\kappa$, and the correction (Eq.~\ref{e:dmu}) specializes to
the derivative of the excess compressibility,
\begin{equation}
\beta \Delta \mu^\text{ex}(\mu,V) = \frac{-1}{2V\kappa }\pd{( \kappa - \beta/\rho )}{\rho}
.\label{e:dmu2}
\end{equation}
This result agrees with Ref.~\citenum{isiep92}, where it was first derived and tested
against the thermodynamic limit for hard-sphere fluids.

\section{ Lattice Gas Model}

  To illustrate the convergence properties of the structure factor approach,
we apply it to a simple one-dimensional, one-component periodic lattice gas
on $L$ sites with nearest-neighbor interaction energy $\beta J$% \equiv x$.
This system is isomorphic to the 1D Ising model.\cite{sfrie17}
The operators, $\hat \rho$, are defined as indicator functions
on the $L$ sites, and all Fourier transforms
become discrete Fourier transforms over $L$.  Otherwise, the theory above
goes through as expected except for a change in the
ideal chemical potential, \cite{jrein00}
\begin{equation}
\pd{\beta\mu^\text{id}}{\rho} = \frac{1}{\rho(1-\rho)}
%Q^{-1}(j) = \delta_{j,0} \pd{\ln (\rho/(1-\rho))}{\rho} - c_j 
.\label{e:c1}
\end{equation}

\begin{figure}
{\centering
\includegraphics[width=3in]{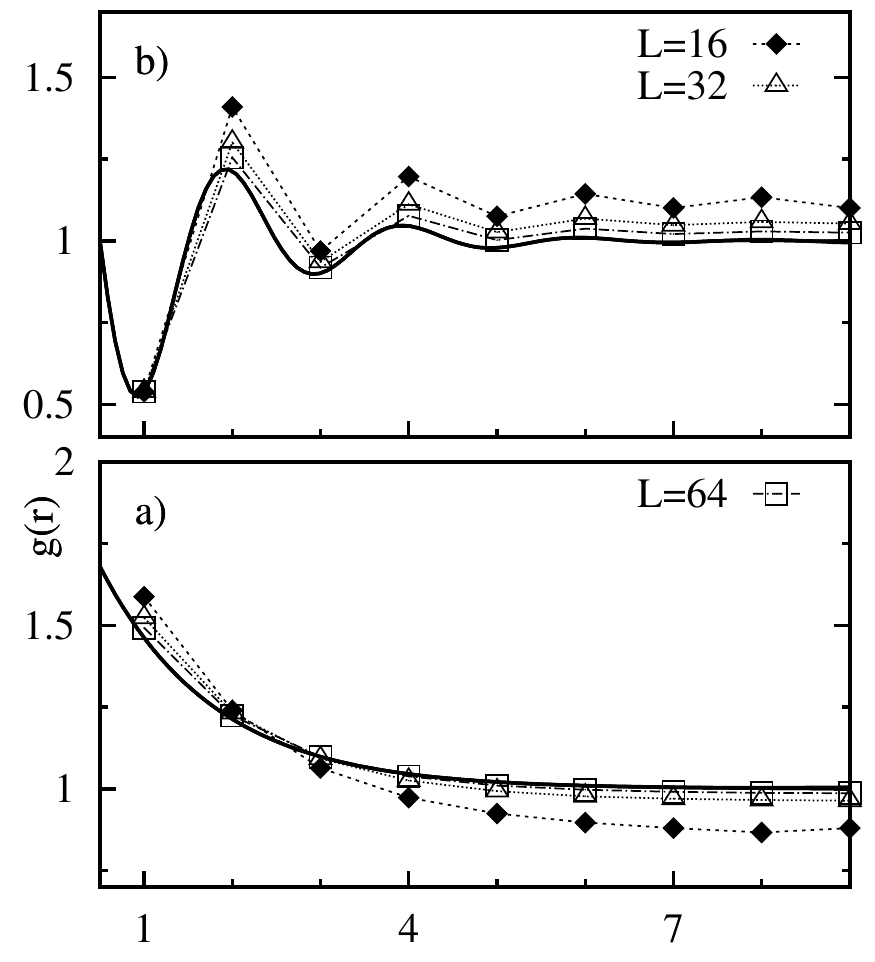}
\caption{Discrete radial distribution functions.
The marked points show canonical, closed simulation data,
while the solid lines show corresponding grand-canonical radial distributions.
The upper panel (a) shows the antiferromagnetic model ($x=-2$), while
the lower panel (b) shows the ferromagnetic $x=2$.
For both plots, the grand-canonical distributions overlap so as to be indistinguishable,
showing strong insensitivity to the cell size, $L$.}
\label{f:gr}}
\end{figure}

  This system, along with several generalizations, were shown to have a strictly nearest-neighbor
direct correlation function in a series of original studies,\cite{cborz87,cteje87} %\cite{cborz87,cteje87,cteje89}
yet it still %has a nontrivial structure factor
poses a nontrivial estimation problem since the correlation length
goes to infinity at strong coupling.
It is ideal for illustrating estimation of the direct correlation function
because the densities and pair distributions can be
computed analytically in the grand-canonical ensemble using the transfer
matrix method.\cite{davis}
We also calculated the canonical ensemble densities and pair distributions
exactly with the use of the absolutely convergent cluster summation technique
of Ref.~\citenum{jvavr01}.
The most important properties of this model depend on
the parameter $y \equiv \pm\exp(-1/L_c)$, where $L_c$ is the correlation length
over which the radial distribution functions decay exponentially.
The complete expression for $y$ in terms of $\beta J$ and $\phi (= \beta\mu)$
is given in the supplementary information.
It can vary from $-1$ in an antiferromagnetic system when
$\beta J \to -\infty$ to $+1$ in a ferromagnetic system when $\beta J \to +\infty$.

\begin{figure}
{\centering
\includegraphics[width=3in]{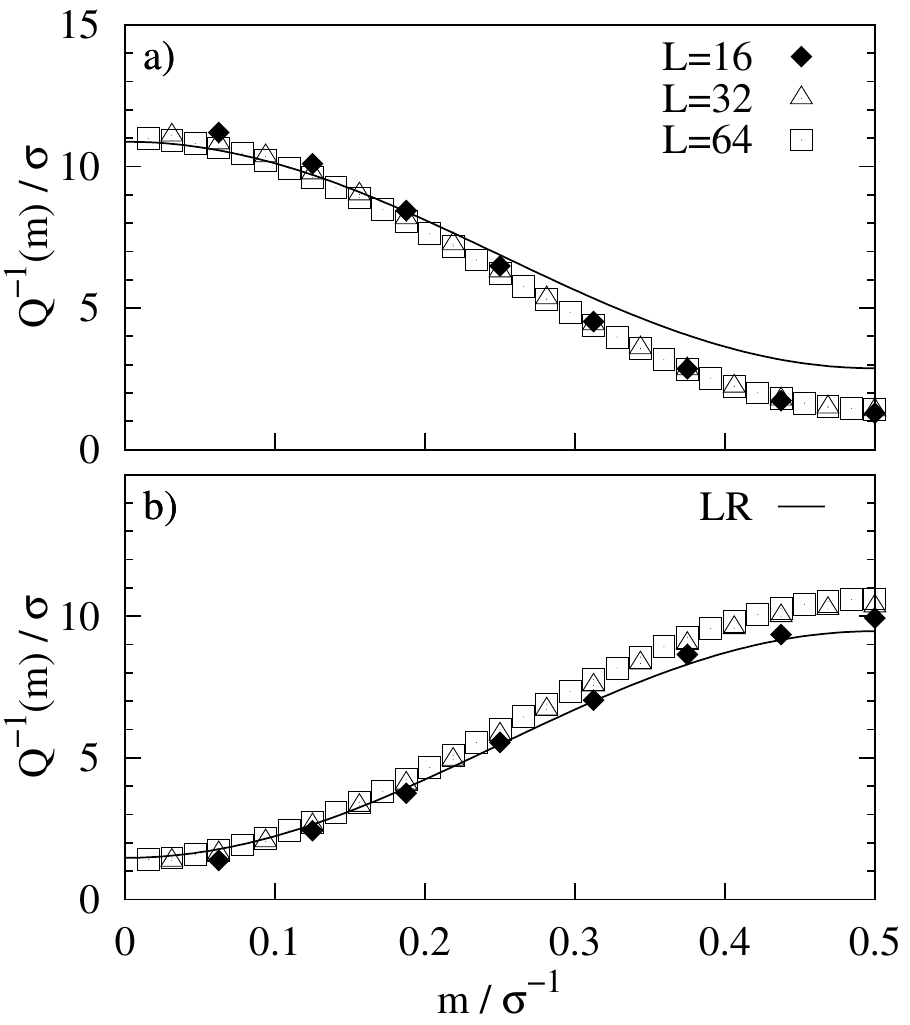}
\caption{Inverse correlation function from Eq.~\ref{e:Qg} without any ensemble corrections.
The marked points show canonical simulation data,
while the smooth lines show the low-density limiting form, $-2\beta J \cos(2\pi m) +\text{const}$.
As in Fig.~\ref{f:gr}, (a) and (b) are antiferromagnetic and ferromagnetic, respectively.}
\label{f:sigma}}
\end{figure}

  Fig.~\ref{f:gr} compares the radial distribution function computed in the canonical
ensemble with the corresponding grand-canonical distributions
(at the chemical potential $\phi = -\beta J$, for which $\avg{n | \mu, L} = L/2$).
The convergence is slow with increasing $L$, even though the canonical correlation functions
appear to flatten out to a constant at large distance.
These shifts in $g(r | n, L)$ by a constant do not affect our analysis,
since they merely change the structure factor at $m=0$ --
which our analysis disregards.  The interesting ensemble effect
shown in this plot is that the $r=1$ point appears to have a different vertical
shift than the large-distance asymptote.
For the antiferromagnetic case (a), the short-range behavior is relatively insensitive to
the ensemble correction.  In contrast (and disregarding the smallest simulation size),
ensemble dependence shows up most strongly
at short range in the ferromagnetic case (b).

  Fig.~\ref{f:sigma} compares the inverse correlation function, $Q^{-1}(m)$ for the
three systems above in the canonical and grand-canonical ensembles.
The function $Q^{-1}$ can be calculated exactly and simplifies
in the limit as $L/L_c \to \infty$,
%it is physically interesting that it can be very closely approximated at large
%coupling, $x$, by,
\begin{equation}
Q^{-1}(m|\mu) = \frac{1}{\rho(1-\rho)} \left(
   \frac{1-y}{1+y} - \frac{2 y}{1-y^2} \left( \cos(2\pi m)-1 \right)
 \right)
\label{e:c2} .
\end{equation}

\begin{figure*}
{\centering
\includegraphics[width=0.9\textwidth]{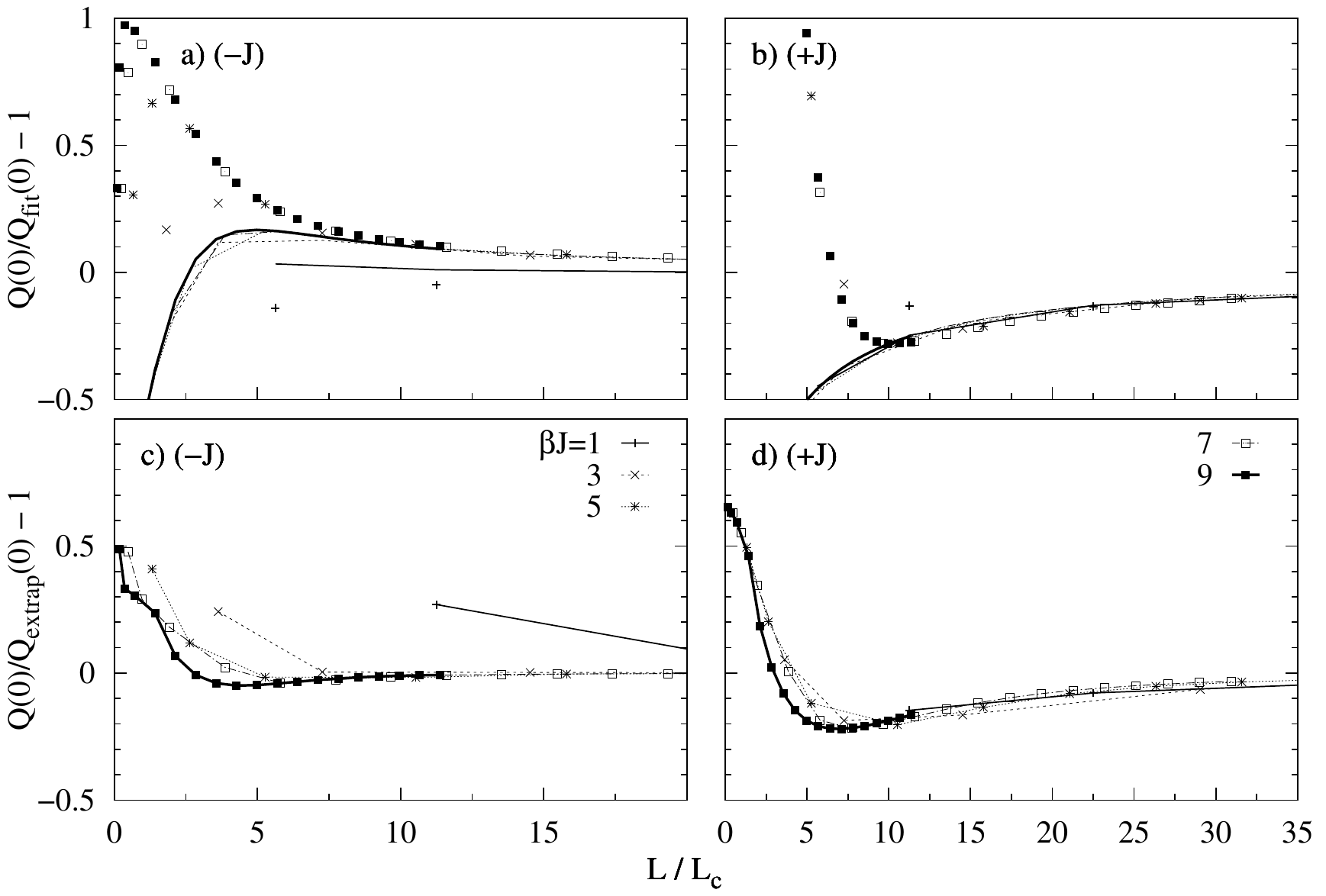}
\caption{Convergence of estimates $Q_\text{fit}^{-1}(m) \sim Q_\infty^{-1}(0) + a m^2$ [3 point fit near $m=0$, (a,b)]
and $Q^{-1}_\text{extrap}(1/L) \sim Q_\infty^{-1}(0) + c/L$ (c,d)
as a function of the ratio of cell size to correlation length.
Each series of points shows estimates derived from $Q(m|n,L)$ (Eq.~\ref{e:Qg})
at the same value of $|\beta J|$.
The lines in (a,b) are second differences of the canonical free energy, showing that
extrapolation yields the canonical KB coefficient.
The antiferromagnetic case ($\beta J < 0$) is left (a,c)
and the ferromagnetic case ($\beta J > 0$) is right (b,d).}\label{f:extrap} }
\end{figure*}

  Fig.~\ref{f:extrap} compares two estimates for the KB coefficients
corresponding to extrapolating to $m=0$ following Eq.~\ref{e:fit}
(panels a,b) and correcting for ensemble dependence, fitting every two successive
lengths to an empirical expression, 
\begin{equation}
Q^{-1}_\text{extrap}(1/L) \sim Q_\infty^{-1}(0) + c/L
,\label{e:empire}
\end{equation}
for panels (c,d).
The latter two plots used Eq.~\ref{e:empire} rather than Eq.~\ref{e:finite}
because it gave much faster convergence.
Several coupling values, $\beta J$, and several simulation sizes are shown,
and fall on a single curve when scaled by their correlation lengths.
The lines in (a,b) plot Eq.~\ref{e:dmuex} for second derivatives (times $L$)
of the Helmholtz free energy for the lattice gas as $n$ is varied.
Its agreement with the extrapolated values
shows that the canonical ensemble $Q^{-1}(0 | n,V)$
converges to \ref{e:dmuex} first, and
then converges to $\partial \beta\mu^\text{ex}/\partial\rho$ as $V$ increases.

  The error visible at $\beta J = -1$ (panel a) gives another important
message for estimating the KB coefficients.
Rather than fit to Eq.~\ref{e:fit}, we should have subtracted off the
known contribution to the direct correlation function, $c_\text{LR} = -2\beta J \cos(2\pi m)$,
in Eq.~\ref{e:OZ} and fit the remainder.
Since the correlation length is very short for this weakly coupled limit,
the resulting estimate would give perfect agreement with the line for $\beta J = -1$
(not shown).

  The lower panels (c,d) show extrapolation to $1/L = 0$ using the
assumption\cite{pkrug13} that $\pd{^2}{\rho^2} Q(m | \mu, L)$
is independent of $L$ in Eq.~\ref{e:finite} and that $Q(1/L) \sim Q(0)$.
Attempts to combine both types of extrapolation, in either order,
resulted in worse convergence (not shown).

  Comparing the ferromagnetic and anti-ferromagnetic cases shows that
the oscillating radial distribution function of the anti-ferromagnetic
system leads to better overall convergence.  This is associated with
better short-range behavior of $g(r)$ (Fig.~\ref{f:gr}a).
The strong ensemble-dependence
of the ferromagnetic $g(r)$ carries over to Fig.~\ref{f:extrap}a, where it appears
as worse agreement at small inverse-distance, $m$.  It is because
of the slow convergence of canonical and grand-canonical chemical potentials
that Fig.~\ref{f:extrap}b shows $Q(0 | n,L)$ is 10\% lower than $Q(0 | \mu, L)$,
even at very large lengths.
Luckily, this issue appears to be limited to the 1D case.

%The 1D Ising model approaches a critical point as $x$ goes to infinity.
%Here, the correlation functions display long-range ordering and the traditional
%KB integral encounters increasing difficulties.

%\begin{table}
%\begin{tabular}{lllll}
%  &  \multicolumn{2}{c}{Antiferromagnetic} & \multicolumn{2}{c}{Ferromagnetic} \\
%  & $L = 10$ & $L > 10$ & $L = 10$ & $L > 10$ \\ \cline{2-5}
%$\partial \mu / \partial c$ & 6.594895 & 6.594885 & 2.426126 & 2.426123 \\
%\end{tabular}
%\end{table}

\section{ Lennard Jones Model}

  This section presents results of applying the formalism to a supercritical Lennard-Jones (LJ) gas.
To mirror the conditions of the lattice model, we fix the density to a near-critical value,
$\rho^* \sigma^3 \approx 0.36$, and set $\beta^{-1} = k_B T = 1.5 \epsilon$.
This is above the critical temperature, $T^* = 1.36 \epsilon/k_B$.
In the lattice model analogy, the coupling constant, $|\beta J|$, would be $\epsilon / k_B (T - T^*)$
if the temperature varied while $\epsilon$ remained constant.
Fluctuations in this system are interesting because they report on
cluster formation processes.\cite{hmart96}
After summarizing target `thermodynamic limit' results computed in the grand-canonical ensemble
and using analytical formulas, we present the scaling behavior in the canonical ensemble.

  The LJ gas has become an indispensable simulation model, for which
analytical approximations many properties are available.
The most important for the present case
are the equation of state, which gives $Q(0)$ via the compressibility, and
the radial distribution function, which gives $Q(m)$ via Eq.~\ref{e:g}.
We compare our simulation results to the MBWR equation of state,\cite{jnico79,jjohn93}
which was parameterized from molecular dynamics and Monte Carlo simulation data,
and includes the effects from tail corrections.  We compare the radial distribution
function with a numerical solution of the Percus-Yevick (PY) closure of the Ornstein-Zernike
equation, computed following the method of Fries and Patey.\cite{pfrie86}

  Our grand-canonical Monte Carlo (GC) simulations were performed with Towhee,\cite{mmart13}
at an excess chemical potential of $\mu^\text{ex} = -1.868 \epsilon$ in
a cubic box of length $10.357\sigma$.  Pair interactions were truncated at 5$\sigma$
after which long-range corrections were added for the energy and pressure.
Simulations were run long enough
to collect 10,000 configurations for analysis.  The number of trial moves between
configurations was 40,000, since the potential energy showed an exponential
autocorrelation function with a decay time of 13,200 trial moves.  Moves were selected
randomly with 25\% probability for particle insertion/deletion and 75\% probability for
translation.
%  Particle numbers, pressures, and energies were saved every 4,000 move attempts.

  Because our GC simulation gave a slightly different density
($0.356 \sigma^{-3}$), the GC properties summarized in Table~\ref{t:GC}
were obtained at $\rho = 0.36\sigma^{-3}$ by differentiating
the partition function estimate,
\begin{align}
\beta P V &= \Theta(\beta\mu) \approx  \Theta(\beta\mu_0) e^{C(\beta\mu - \beta\mu_0)} \\
C(\lambda | \mu, L) &\equiv \ln \sum_{n=0}^\infty  f_n(\mu,L) e^{\lambda n} \label{e:C}
,
\end{align}
where $f_n$ is the frequency of observing $n$ particles in the GC simulation.
Eq.~\ref{e:C} is a maximum likelihood estimate appropriate for a large number of samples,
$S$, when the probability for observing a histogram, $f$, is given by Sanov's theorem.
We computed the estimation error using the saddle-point approximation
(taking the second derivative of the relative entropy) to yield the covariance,
\begin{equation}
\avg{\delta C(\lambda)\delta C(\lambda')} S = e^{C(\lambda + \lambda') - C(\lambda) - C(\lambda')} - 1
.
\end{equation}

\begin{table}
\begin{tabular}{llllll}
  L/$\sigma$ & $\mu^\text{ex} / \epsilon$ & $\beta P/\rho$ & $Q^{-1}(0)/\sigma^3$
   & $L_c / \sigma$ & $\pd{Q^{-1}(0)}{(\beta\mu)}/\sigma^3$ \\ % & $\frac{Q(0)}{2} \pd{^2 Q(0)}{\rho^2}$ \\ 
  \hline \hline
6.525  & -1.887(3)  & 0.4597(11) & 1.89$^*$ & 0.70$^*$ &  \\
8.434  & -1.886(3)  & 0.4670(7)  & 1.81 & 0.97 &  \\
10.36  & -1.884(3)  & 0.4697(5)  & 1.53 & 1.16 &  \\
12.65  & -1.882(3)  & 0.4720(4)  & 1.32 & 1.36 &  \\
16.44  & -1.882(3)  & 0.4730(3)  & 1.24 & 1.49 &  \\
19.68  &                  & 0.47367    & 1.27 & 1.50 &  \\
25.30  &                  & 0.47312    & 1.25 & 1.55 &  \\
\hline
% -1.88709 (insertion, no reweighting on n)
% PY fit of Q(m) gives L_c = 3.65, but integrals of c(r) give 1.77
GC      &  -1.878(3) & 0.47591  & 1.084(9) & 1.50 & 4.0(3) \\% & -20(11) \\
PY       &                 &                & 1.28        & 1.77 & 4.0724 \\% & -21 \\
MBWR & -1.854     & 0.48218  & 1.1330    &         & 3.3978 \\%& -19
\hline
\end{tabular}\caption{ Comparison of excess chemical potential, pressure, and its derivatives from canonical simulations (top 7 rows), and grand-canonical and equation of state calculations (bottom rows).
Numbers in parentheses display the uncertainty in the last digit.
$^*$Only the first two points were used in the extrapolation to $m=0$
for the smallest simulation size.}\label{t:GC}
\end{table}

  The MBWR equation of state is fairly accurate near the state point studied here,
but predicts $Q^{-1}(0)$ that is too low by $0.06 \sigma^{3}$ and $\partial Q^{-1}(0)/\partial (\beta\mu)$
that is too low by $0.7 \sigma^{-3}$.  Although its parameterization\cite{jjohn93} did not include the particular
state point studied here, it can be noted that the MBWR slightly over-estimates
$\beta P / \rho$ in this region.  That quantity reaches a minimum near the present simulation
conditions, and it is the shape of the minimum which the derivatives in Table~\ref{t:GC} report on.
We estimated the finite-size correction for $Q(m)$ in Eq.~\ref{e:finite},
from GC, PY and MBWR methods and all three agree on $\frac{Q(0)}{2} \pd{^2 Q(0)}{\rho^2} = -20$,
although the error in the GC estimate is 50\%.

%  For high densities, the Weeks-Chandler-Anderson approximation\cite{jweek71} does
%well reproducing the radial distribution function.  However, it has noticeable
%errors at low densities -- around $\rho^* \sigma^3 = 0.36$ in particular.

  Canonical simulations were performed with Towhee for $L=6.525\sigma$
to $L=16.44\sigma$.  The smallest two simulations used pair interaction
cutoffs of $3.25\sigma$ and 4.2$\sigma$, respectively.  All larger cell sizes
used a cutoff of 5$\sigma$.  All results include long-range corrections for the
energy and pressure.  The largest two simulations were run with LAMMPS
using a timestep of $0.003$ LJ units and collecting 10$^4$ samples
-- each separated in time by 10$^4$ dynamics steps.
The $Q(0)$ and $L_c$ values reported in Table~\ref{t:GC}
were fit to the smallest 3 $m$-points following Eq.~\ref{e:fit}.
For the $L=6.525\sigma$ simulation, only the smallest 2 points
were used in the fit.

\begin{figure}
\includegraphics[width=0.45\textwidth]{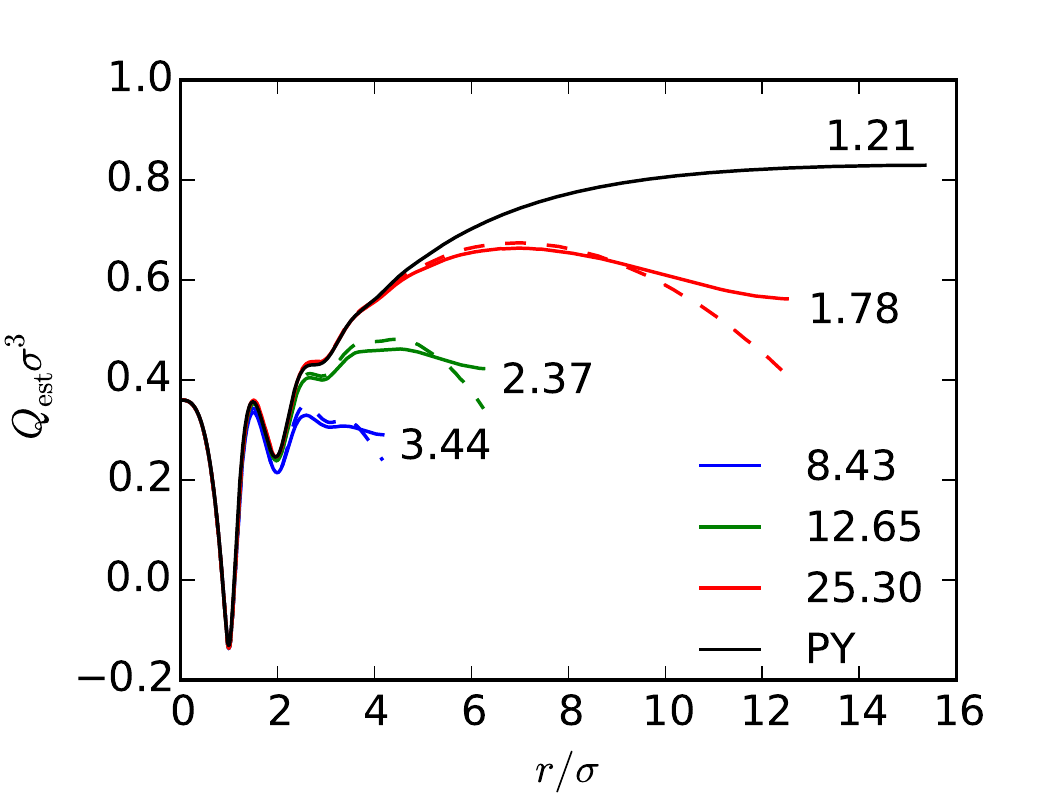}
\caption{Estimates of $Q(0)$ using the radial distribution
function integral (Eq.~\ref{e:Gint}) from different simulation sizes.
The value at the origin is the ideal gas contribution, $\rho = 0.36\sigma^{-3}$,
while the final estimate must be positive.
Dotted lines show the partial integral without the correction of Ref.~\citenum{pkrug13}.
The inverse of the final values (labeled on the plot) should be compared to Tbl.~\ref{t:GC}.
The PY number is smaller than 1.28 due to the integral cutoff.}\label{f:Gint}
\end{figure}

  Figure~\ref{f:Gint} shows a running integral of the radial distribution function
from several of the simulations.  The solid lines use the correction
proposed by Kr\"{u}ger et. al.,\cite{pkrug13}
\begin{equation}
Q_\text{est} = \rho + \rho^2 \int_0^{L/2} 4\pi r^2 (g(r)-1) \left(1 - (\tfrac{r}{L/2})^3 \right)
,\label{e:Gint}
\end{equation}
where $L$ is the length of one side of the simulation cell.
The dotted lines were not corrected (setting the $1 - 2r/L$
term to $1$ in Eq.~\ref{e:Gint}).
The figure also shows that the PY closure yields a reasonable estimate of the RDF
up to $4\sigma$, while the remainder is not possible to judge based
on our simulation sizes.  The long correlation length of this near-critical point
system makes for a challenging test case.

\begin{figure}
\includegraphics[width=0.45\textwidth]{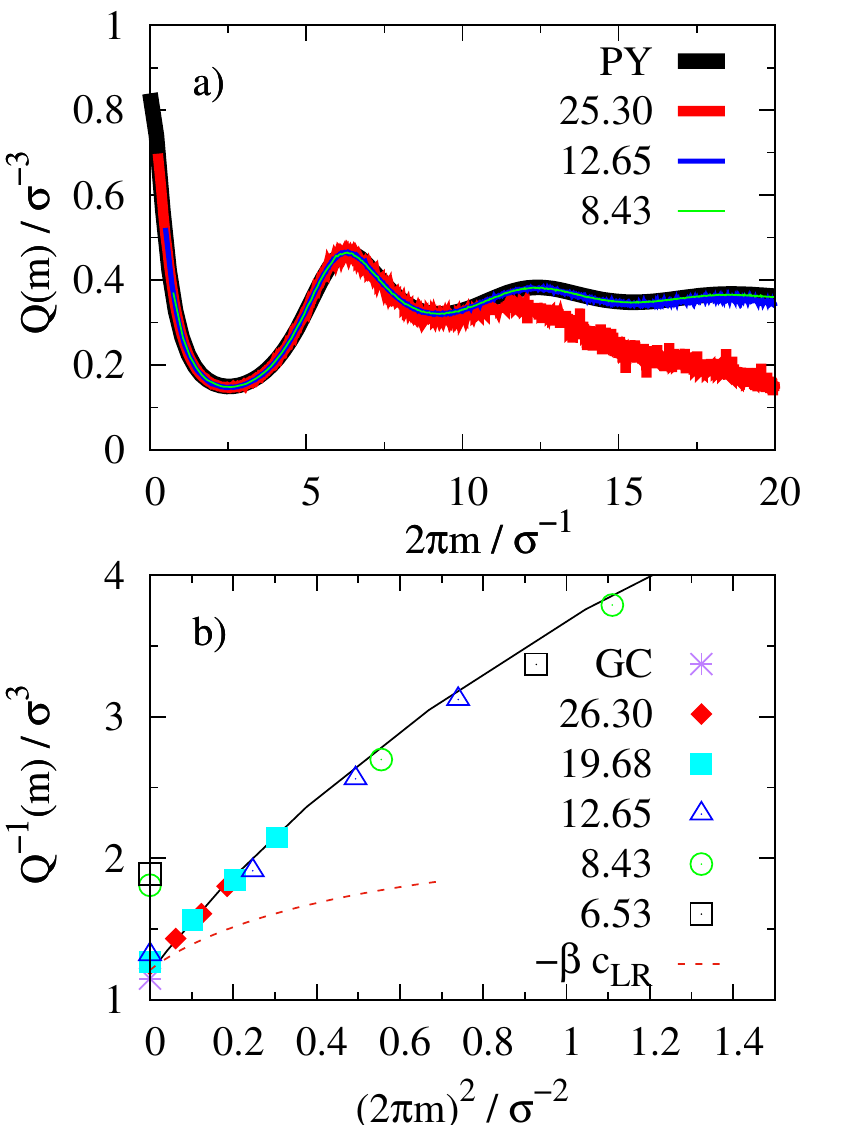}
\caption{Estimates of $Q(m)$ from the Lennard-Jones system for the same simulations
shown in Fig.~\ref{f:Gint}.  The $L = 25.30\sigma$ line shows artificial damping
at large $m$ caused by the smoothing spline FFT method on our fixed 128$^3$ grid.
Otherwise, all the lines in (a) overlap one another.
In (b), only the smallest three $k$-points from each simulation are shown, and the horizontal axis is scaled
to show $(2\pi m)^2$, anticipating the role of the correlation length (Eq.~\ref{e:clen}).
The dotted line plots Eq.~\ref{e:uessm} plus a constant using $\eta = 1$.}\label{f:LJ}
\end{figure}

  Fig.~\ref{f:LJ}a plots the structure factor, $Q$, using spectroscopy units, $k = 2\pi m$.
It should be compared with Figs.~3 and~4 of Ref.~\citenum{jweek71},
where the relationship is $\rho^2 \tilde h(k) = Q(k/2\pi) - \rho$.
That reference shows the shape of $Q(m)$ near $m=0$ was very flat, and
easy to extrapolate.  The present case is much more difficult because it is near
the critical point and a very large peak has appeared near $m=0$.

  Fig.~\ref{f:LJ}b shows the inverse of Fig.~\ref{f:LJ}a near the origin.
It is the analogue of Fig.~\ref{f:sigma} for the lattice gas.
For clarity, only the first three points with smallest $m$ from each simulation are shown.
These particular $m$-points correspond to averages over plane wave shaped density fluctuations
oriented along faces, edges, and corners of the cubic unit cell.
All five of the canonical simulation points shown at $m=0$ are extrapolations
from these points.  Their numerical values are in Table~\ref{t:GC}.

  Fig.~\ref{f:LJ}b also shows two other important points.
First, $Q(m)$ computed from the largest simulation size
is artificially noisy and scaled downward at high wavenumbers.
The scale artifact is due to the inaccuracy of B-spline smoothing
past $K_\text{max}/2$, where $K_\text{max} = 128/L$ is set by
our numerical grid, which contained 128$^3$ points for every $L$.
A similar artifact also appears for the $L=12.65$ simulation just
past the right boundary of Fig.~\ref{f:LJ}b.  The noise is
caused by the discrete nature of the 128$^3$ grid points, which appear
at irregularly spaced distances from the origin.  It could be eliminated
by resamping the data to uniform intervals in $|m|$.

  Second, the long-range portion of the LJ potential can be Fourier-transformed
according to Ref.~\citenum{uessm95} to give,
\begin{equation}
c_\text{LR}(x = m/\eta) = \frac{4\epsilon}{3} \left[
(1-2x^2) e^{-x^2} + 2 x^3 \sqrt{\pi} \erfc(x)
\right]
,\label{e:uessm}
\end{equation}
where $\eta$ is an inverse distance corresponding to a
division between short-range and long-range forces.
Recognizing that this term makes an additive contribution to Eq.~\ref{e:OZ}
can accelerate the convergence of the fit in Eq.~\ref{e:fit}.
Indeed, we find that subtracting this term from $Q^{-1}$
makes the first few points much more linear.  The result can be visualized
from Fig.~\ref{f:LJ}b as the distance between the red and black lines.
However, to keep our discussion simple, we have not included it in our estimates.

  Similar to the antiferromagnetic lattice gas in Fig.~\ref{f:extrap}a,
the estimates of $Q(0)$ are larger than the grand-canonical value,
and decrease with increasing cell size.  The former figure showed that
the extrapolated points, $Q(0)$, quickly converged to second differences of
the Helmholtz free energy (Eq.~\ref{e:dmuex}).
We cannot test this numerically here, since the MD estimates of $\beta\mu^\text{ex}$
at finite volumes have uncertainties on the order of the volume.

  Instead, we have tested the finite-size corrections to $\beta\mu^\text{ex}$
in Eq.~\ref{e:dmu}.  Using the value of $Q^{-1}$ and its derivative
from Table~\ref{t:GC} gives $\Delta \beta \mu^\text{ex} = 4.18\sigma^3 / V$.
This is a good fit to the size-dependence of $\mu^\text{ex}$ seen in simulations.
In the limit as $m \to 0$, the finite-size corrections to $Q$
and to $\beta\mu^\text{ex}$ are related by a density derivative.
Computing the finite-size correction to $Q(0)$ from Eq.~\ref{e:finite}
gives $\lim_{m\to 0} \Delta Q(m) = -20 / V$, which agrees with
the density derivative of Eq.~\ref{e:dmu}.
The simulation points do not allow a single estimation of the
slope near $m=0$ for the largest simulations,
but the slope of the $c_0$ values vs. $1/V$ appears to be +200 rather
than -20 for simulations at intermediate cell-sizes.

\begin{figure}
\includegraphics[width=0.45\textwidth]{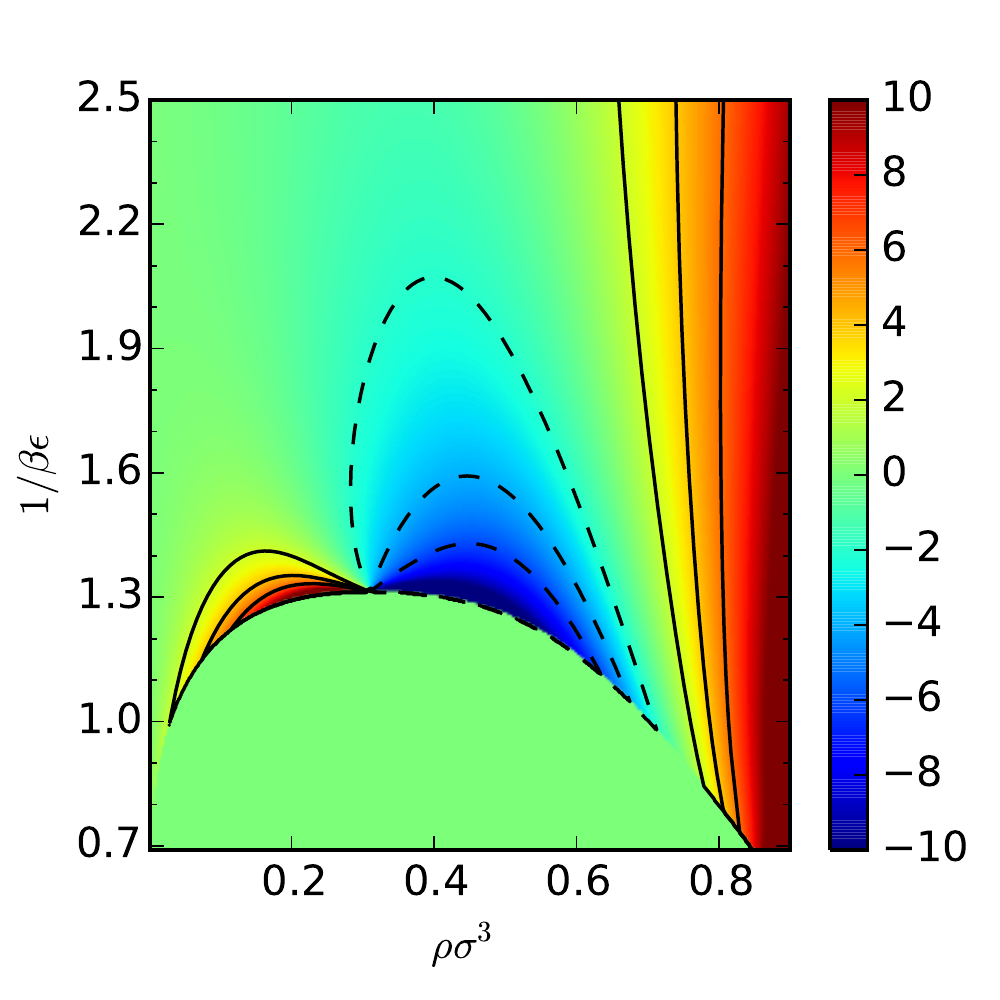}
\caption{Contour plot of the $1/V$ coefficient in the finite-size correction to the free energy,
$\partial(\kappa - \beta/\rho)/\partial\rho / 2\kappa$ (Eq.~\ref{e:dmu2})
as calculated using the MBWR equation of state.
Contours are shown every 2$\sigma^3$, while variations above
$\pm 10 \sigma^3$ were colored at the max/min values.
The axes are LJ temperature and density.  Values in the liquid-vapor coexistence
region (bottom, under the saturation curve) have been set to zero for clarity.}\label{f:finite}
\end{figure}

  Because finite size corrections are relatively more important for canonical than grand-canonical
simulations, we have plotted the finite size correction to the free energy
predicted by the MBWR equation of state in Fig.~\ref{f:finite}.
The result reassures us that the chemical potential calculated by
test particle insertion has negligible volume dependence in both gas and liquid
states.% above more than 20 molecules or so.
Near the solidification line, test particle insertion should be expected to slightly overestimate the
excess chemical potential, while we get an underestimation for high-temperature
liquids near the critical point.

\section{ Conclusions}

%  Kirkwood-Buff theory derives several thermodynamic quantities related to
%$\tilde \phi_\alpha(0) = \beta\mu_\alpha$.  When the potential is short-ranged
%so that $Q(m)$ is continuous near the origin, then by continuity of $\tilde \phi(m)$,
%$\Sigma^{-1}(m)$ near $m=0$ can be used in the same expressions
%to find the partial molar volumes, isothermal compressibility,
%and chemical potentials in the constant-pressure and ensemble
%as well as the dependence of the osmotic pressure on solute interactions.
%It is an interesting open question whether $S_\alpha(m)$ may be
%treated as quasi-particles (as in the theory of spin waves)
%to extend to them the equilibrium concepts of partial molar quantities.

  Kirkwood-Buff integrals are notoriously difficult to calculate from canonical, closed
simulations.  We have shown that this problem
can be helpfully mapped onto the problem of estimating the direct correlation
function of Ornstein-Zernike theory in a canonical ensemble.
All earlier work used truncated real-space estimates, for which the process of extrapolating
to infinite truncation radius, $R$, is ill-posed.\cite{nmatu96,pgang13}
Our exact theory for the canonical ensemble uses the entire simulation volume at once,
is manifestly invariant to shifts of the radial distribution function by a constant, and replaces
the problem of extrapolating to infinite $R$ with the well-posed problem
of extrapolating to zero $m$ within a finite simulation volume.

  The density-dependence of the radial
distribution function becomes important when
estimating the thermodynamic limit.\cite{jlebo67,jsala96}
These higher-order derivatives are very difficult to estimate accurately
from simulation data, and it is simpler to test for finite size effects
by scaling up the system.\cite{pkrug13}
This work showed that %the KB coefficients from canonical simulations converge
%to differences in Helmholtz free energies, and thus
the principle difficulty in reaching
the thermodynamic limit is the size-dependence of the canonical ensemble itself.
We then provided a simple, general formula for estimating finite-size corrections
of the excess chemical potential from KB coefficients.
These corrections are small for the LJ fluid.

  Our data in Fig.~\ref{f:LJ} show excellent agreement for the
entire correlation function at low wave-vectors in reciprocal space.
Higher wave-vectors require smaller grid-spacing to avoid numerical
artifacts, though.  Fits using Eq.~\ref{e:fit} provide an estimation of both
the KB coefficient and the correlation length which are more accurate than
radial-distribution function integrals and fit well into the context of integral theories
of solution.  We have shown that the theory is robust by
treating difficult cases with long-range correlations.  These can be identified
in simulation work from the shape of $Q(m)$ near $m=0$.

  We have shown that the zero-frequency limit of the canonical KB
theory gives second derivatives of the Helmholtz free energy for simulation sizes
larger than about 5 times the correlation length.
Further, this theory has several nice properties.
In the low density limit, the direct correlation function can be predicted
from the long-range form of the interaction energy function.
This leads to even better extrapolation to zero frequency.
Further information on the applicability of this process is given in the supplementary material.
Alternatively, the direct correlation function can be computed from simulation data
and used to test common assumptions for OZ closure relations.\cite{pdt6}
A companion paper uses this process to compute the spatial
dielectric response of water.\cite{droge18a}
This use of $Q$ to estimate pair interaction energies
was first proposed by Madden and Rice~\cite{wmadd80}.

\section*{ Supplementary Material}
  The supplementary material includes derivations and extended discussion of several formulas from
the main text.  It provides a definition of the finite-volume radial distribution, free from cutoffs,
and consistent with Eq.~\ref{e:g}.
It also provides explicit expressions for the other derived quantities of Kirkwood-Buff
theory in terms of the matrix $Q_{\alpha\gamma}(0|\mu)$.
It derives the ensemble correction to the free energy (Eq.~\ref{e:dmu}).
Then it gives computable expressions for the correlation function of the finite-size
Ising model and an approximate expression for the correlation length of the Lennard-Jones fluid.
Both correlation lengths are plotted as a function of state point.
Those results provide additional intuitive insight on the long-range correction proposed above.

\section*{ Acknowledgments}
  This work was supported by the USF Research Foundation and NSF MRI CHE-1531590.

%\bibliographystyle{unsrt}
%\bibliography{stat}

\renewcommand{\theequation}{S\arabic{equation}}
\renewcommand{\thefigure}{S\arabic{figure}}
\renewcommand{\bibnumfmt}[1]{[S#1]}
\renewcommand{\citenumfont}[1]{S#1}
\setcounter{section}{0}
\setcounter{equation}{0}
\setcounter{figure}{0}

\section{ Ensemble Dependence}

  Explicit expressions for the radial distribution functions
help to show how ensemble dependence arises in the KB integral.
Eq.~1 of the main text can be written as an average over
canonical simulations,
\begin{align}
\avg{\hat \rho_\alpha(r) \hat \rho_\gamma(0) | \mu, L} &= \sum_{n} P(n | \mu, L)
\avg{\hat \rho_\alpha(r) \hat \rho_\gamma(0) | n, L} \notag \\
\avg{\hat \rho_\alpha(r) \hat \rho_\gamma(0) | n, L} &= \frac{n_\gamma}{V} \sum_{j=1}^{n_\alpha} P(r_{\alpha,j} = r | r_{\gamma,1} = 0, n, L) \label{e:n2}
\end{align}
The conditional probability in the last line gives the probability that the $j^\text{th}$
molecule of type $\alpha$ has its center of mass at $r$ given that
the first molecule of type $\gamma$ is centered at the origin.
When $\alpha = \gamma$, this has the effect of decomposing the sum over
$j$ into $j=1$, where we know $r_j$ is already
at the origin, and $j \ne 1$, where $j$ is distinct
from the molecule at the origin.
This gives a delta function at the origin plus $n_\alpha - 1$ times
the pair distribution function for fixed $n$.  Explicitly,
\begin{equation}
P(r = r_{\alpha,j} | r_{\gamma,1} = 0, n, L) = \begin{cases}
\delta(r), & \alpha = \gamma, j = 1 \\
g_{\alpha\gamma}(r | n, L)/V, & \text{o.w.}
\end{cases} \label{e:Pr2}
\end{equation}
and
\begin{equation}
\avg{\hat \rho_\alpha(r) \hat \rho_\gamma(0) | n, L}
=  \frac{n_\gamma}{V} \left( \frac{n_\alpha - \delta_{\alpha\gamma}}{V} g_{\alpha\gamma}(r | n, L)
+ \delta_{\alpha\gamma}\delta(r) \right)
\label{e:gV}
\end{equation}

  The identification of $g(r | n, L)$ in Eq.~\ref{e:gV} agrees with its method of
computation from canonical ensemble simulation data by estimating the conditional probability,
$P(r = r_{\alpha,j} | r_{\gamma,1} = 0, n, L)$.
Of course, when the two particles are independent, the conditional
probability is $1/V$, and $g(r | n, L)$ is 1.
However, even for real molecules with excluded volume, $g(r | n, L)$ always
integrates to the volume, $V$.  This makes the integral of $g(r | n, L)$
uninformative for number fluctuations -- since there are none in
a canonical simulation.
Explicitly, if we insert Eqns.~\ref{e:gV} and~\ref{e:n2} into Eq.~1
and integrate, then the $g(r | n, L)$ term drops out and the
entire Kirkwood-Buff integral
is due to the ensemble correction.

  For completeness, we re-state the derived quantities of Kirkwood-Buff
theory in terms of the $\nu\times\nu$ matrix ${\bf Q} = [Q_{\alpha\gamma}(0|\mu,L)]$ and using
the vector of $\nu$ densities, $\rho$, below.
\begin{align}
\kappa/\beta &= 1 / \rho^T {\bf Q}^{-1} \rho \\
\bar V &= \frac{\kappa}{\beta} {\bf Q}^{-1} \rho \\
\beta V \left[ \pd{\mu_\alpha}{n_\gamma} \right]_{n,P,T} &= {\bf Q}^{-1} - \frac{\beta}{\kappa} \bar V \bar V^T \\
\left[ \pd{\Pi}{\rho_\alpha} \right]_{\mu_1,\rho_{\gamma \ne 1},T} &= {\bf Q_2}^{-1} \rho_2
\end{align}
These formulas express the isothermal compressibility, vector
of partial molar volumes, matrix of constant-pressure chemical potential derivatives,
and vector of $\nu-1$ osmotic pressure derivatives (respectively).
The derivative of the osmotic pressure requires a special notation, ${\bf Q_2}$, for the
$\nu-1\times\nu-1$ sub-matrix of ${\bf Q}$ discarding the first row and column, and a
similar notation, $\rho_2$ for the $\nu-1$ component vector of solute densities.
Note that the entries are discarded from $\bf Q$ before inverting.
The equation for $\partial \mu/\partial \rho$ at constant temperature and
pressure requires using the relation $dn_1 \bar V_1 + dn_2 \bar V_2 = 0$
for a constant pressure variation of $n_1$ at constant (arbitrary) volume,
which does not generalize beyond 2-components.

\section{ Ensemble Correction for the Excess Chemical Potential}

\begin{widetext}
  The exponent of the excess chemical potential
can be written in the grand-canonical ensemble as,
\begin{equation}
e^{-\bmuex{\alpha}} = e^{\beta(\mu^\text{id}_\alpha - \mu_\alpha)} \equiv Z
,
\end{equation}
which we call $Z$ for compactness.
The ensemble correction for $Z$ requires alternating
derivatives,[15]%\cite{jlebo67}
\begin{equation}
Z(\mu) = Z(n) + \frac{V}{2 V} \sum_\gamma
\pd{}{(\beta\mu_\gamma)} \pd{}{n_\gamma} Z(\mu)
, \label{e:corr}
\end{equation}
evaluating the first at constant chemical potentials ($\mu_\alpha$)
and the second at constant densities ($\mu_\alpha^\text{id}$).
The volume is constant in both, and can be used to transform
$n_\gamma / V = \rho_\gamma$.
The first derivative is,
\begin{equation}
\pd{}{\rho_\gamma} Z(\mu) = Z(\mu) \left(
     \pd{\bmuid{\alpha}}{\rho_\gamma}
    - Q^{-1}_{\alpha\gamma}(0) \right)
.
\end{equation}
The next step requires the other derivative,
\begin{equation}
\pd{}{(\beta\mu_\gamma)} Z(\mu) = Z(\mu) \left(
  \pd{\bmuid{\alpha}}{\rho_\alpha} Q_{\alpha\gamma}(0)
  - \delta_{\alpha\gamma} \right)
,
\end{equation}
which uses the fact that $\mu^\text{id}_\alpha$ does not
depend on any densities other than $\rho_\alpha$.
Using these in Eq.~\ref{e:corr} yields, (to order $1/V$)
\begin{equation}
\frac{Z(\mu)}{Z(n)} = 1 + \frac{1}{2V} \left(
  - 2 \pd{\bmuid{\alpha}}{\rho_\alpha}
  + Q_{\alpha\alpha}(0) \left[
            \left(\pd{\bmuid{\alpha}}{\rho_\alpha} \right)^2
            + \pd{^2\bmuid{\alpha}}{\rho_\alpha^2} \right]
  + Q^{-1}_{\alpha\alpha}(0)
  - \sum_\gamma \pd{Q^{-1}_{\alpha\gamma}}{(\beta\mu_\gamma)}
\right)
.\label{e:corr2}
\end{equation}
\end{widetext}
Since $\partial\bmuid{\alpha} / \partial \rho_\alpha = 1/\rho_\alpha$,
the $Q_{\alpha\alpha}(0)$ term cancels.
The final ensemble correction given in the main text is the first term
in the expansion of $-\ln Z(\mu)/Z(n)$.
Since $Q^{-1}$ is naturally a function of $\rho$, while its derivative in
Eq.~\ref{e:corr2} is taken at constant $\rho$,
it is useful to change the independent
variables from $\mu$ to $\rho$,
\begin{equation}
\pd{Q^{-1}_{\alpha\gamma}}{(\beta\mu_\gamma)}
  = \sum_\delta Q_{\gamma\delta}(0) \pd{Q^{-1}_{\alpha\gamma}}{\rho_\delta}
.
\end{equation}
Another alternate formula is,
\begin{equation}
\pd{Q^{-1}_{\alpha\gamma}}{(\beta\mu_\gamma)}
 = -\sum_{ij} Q^{-1}_{\alpha i} \pd{Q_{ij}}{(\beta\mu_\gamma)} Q^{-1}_{j\gamma}
.
\end{equation}

\section{ Ising Model Distribution Function}

  The lattice gas in the main text has potential energy function,
\begin{equation}
\beta H(\hat \rho | L) = -x \sum_{j=1}^L \hat\rho_{j-1} \hat\rho_j
,
\end{equation}
where we have defined $\beta J \equiv x$ and used
periodic boundary conditions, $\hat\rho_L = \hat\rho_0$.
To compute powers of the transfer matrix,
\begin{equation}
T \equiv \begin{bmatrix}
e^{\phi + x} & e^{\phi/2} \\
e^{\phi/2} & 1
\end{bmatrix}
,
\end{equation}
we write it as $T = \alpha I + r_x \sigma_x + r_z \sigma_z$,
where $\sigma_x$ and $\sigma_z$ are $2\times 2$ Pauli
spin matrices and
\begin{equation}
\alpha \equiv (e^{\phi + x} + 1)/2, \quad
r \equiv [ e^{\phi/2}, 0, (e^{\phi + x}-1)/2 ]^T
.
\end{equation}
We also define the scale parameter,
\begin{equation}
y \equiv \frac{\alpha - |r|}{\alpha + |r|}.
\end{equation}

  The partition function can then be written as,
\begin{align}
Z(\phi, L) &= \Tr{ T^L } = (\alpha + |r|)^L (1 + y^L) \notag \\
\frac{T^m}{(\alpha + |r|)^m/2} &= (1 + y^m) I + (1 - y^m) (r_x \sigma_x + r_z \sigma_z)/|r|
, \\
\intertext{and it is easy to find expectation values
of $\hat \sigma_z(j)$,}
\avg{\hat \sigma_z(0)} &= \Tr{ T^L \sigma_z }/Z = \left( \frac{1 - y^L}{1 + y^L} \right) \frac{r_z}{|r|} \\
\avg{\hat \sigma_z(0) \hat \sigma_z(j)} &= \Tr{T^{L-j}\sigma_z T^j\sigma_z}/Z \notag \\
  &= \frac{r_z^2}{|r|^2}
+ \frac{r_x^2}{|r|^2} \frac{ y^j + y^{L-j}}{1 + y^L} \label{e:gj}
.
\end{align}
These can be transformed into average densities and radial
distribution functions by substituting,
$\hat \sigma_z(j) = 2 \hat \rho(j) - I$.

\begin{figure}
\includegraphics[width=0.45\textwidth]{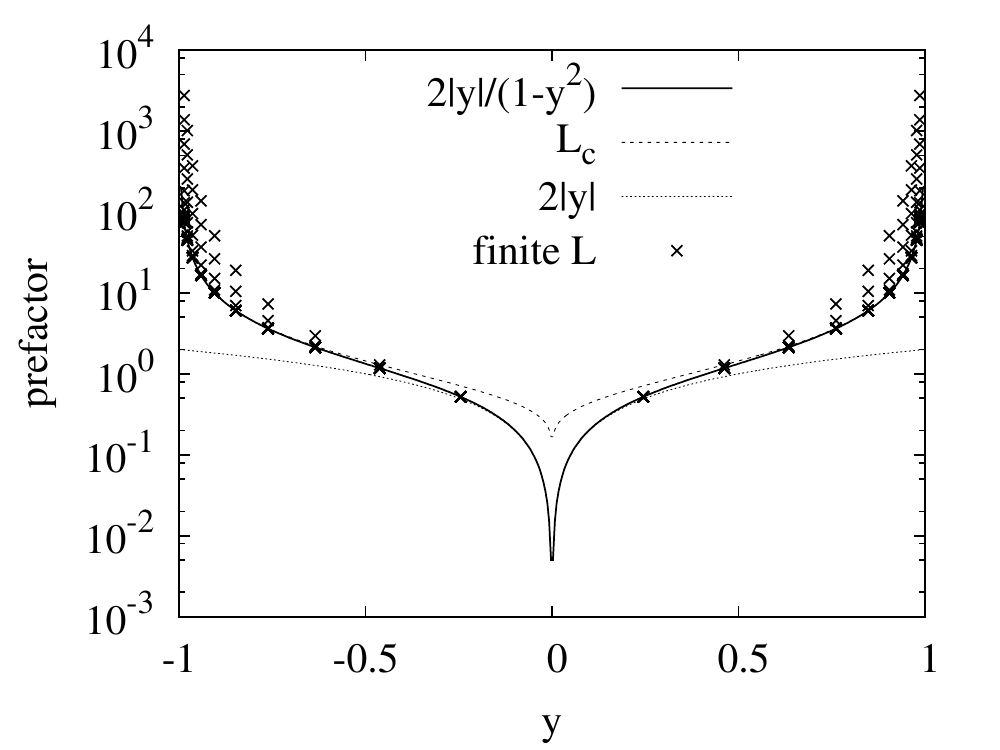}
\caption{Scale of the cosine term in Eq.~18 compared with
the correlation length, $L_c$, and the interaction energy, $y/\rho(1-\rho) \sim x$ as $x\to 0$.
Points mark the scale observed in closed simulations from $L=10$ to $L=512$.
The largest $L$ values approach the line most closely.}\label{f:gc}
\end{figure}

  Fig.~\ref{f:gc} shows the prefactor of the cosine-term in
$Q^{-1}(m)$, $2y/(1-y^2)$.
Since the prefactor dictates the correlation length estimated
by Eq.~16 of the main text, it is important to compare
to the analytical $L_c$ in this model.
In the high temperature limit, $|x| \to 0$, and
the prefactor simplifies to $2x \rho (1-\rho)$.
At large coupling, the prefactor approaches
the analytical correlation length, $L_c$.
Both limits are shown in the figure.

  The finite-size effect on $L_c$ was calculated by estimating
the prefactor for canonical, closed systems.
This estimation is very nearly exact, since $Q^{-1}$ for closed
systems maintains its cosine shape, even at high coupling.
Those estimates are shown by crosses in the figure.
All these estimates show a larger effective correlation
at finite sizes.  Our simulation lengths were spaced
logarithmically in $L$, so the figure suggests
that the prefactor scales linearly with $L/L_c$.

\section{ Correlation Length of the Lennard-Jones System}

  The main text suggests estimating $Q(m)$ near $m=0$
by subtracting a long-range estimate for
the direct correlation function,
\begin{equation}
-\tilde c_{LR}(m) = \int e^{-2\pi i m\cdot r} u_{LR}(r) dr
.\label{e:mod}
\end{equation}
In a companion paper,[30]%\cite{droge18}
we show that this estimate is near quantitative accuracy for
ionic systems away from any phase transitions.

%  Since the correlation length is a structural property, it
%cannot be directly calculated from the equation of state.

  To apply this ansatz here, we modify the WCA approximation
by using a fictitious reference system.
WCA showed that the hard-sphere reference system gets
the high-wavenumber behavior of $\tilde c(m)$ right.
Since $\tilde c_{LR}(m)$ is small at high wavenumbers,
we simply add it to the hard-sphere solution in the
mean spherical approximation (MSA),
\begin{equation}
c_A(r) = c_d(r; \rho\sigma^3) - \beta \epsilon u_{LR}(r)
.\label{e:modA}
\end{equation}

  In variance with the main text, we let $u_\text{LR}$
be the long-range part of the WCA potential,
\begin{equation}
u_{LR}(r) = \begin{cases}
-1, & r < 2^{1/6} \\
4(r^{-12} - r^{-6}), & r \ge 2^{1/6}
\end{cases} ,
\end{equation}
and $r$ is in units of $\sigma$.
The function $c_d$ is the cubic polynomial on $[0,d)$
given by the MSA solution
of the hard sphere system for a diameter chosen
by a consistency condition on $Q(0)$.
%\begin{equation}
%\begin{split}
%d &\simeq \frac{0.3837 + 1.068\beta\epsilon}{0.4293+\beta\epsilon}\Big[1 + \\
%&\quad \frac{-\tfrac{17}{4}\eta_W + 1.362\eta_W^2 -0.8751\eta_W^3}{(1-\eta_W)^2 (210.31 + 404.6\beta%\epsilon)} \Big]
%\end{split}
%\end{equation}
%and density $\rho$.\cite{VW}
%The packing fractions are defined as $\eta \equiv \pi\rho d^3/6$ and $\eta_W \equiv \eta - \eta^2/16$.
We then multiply the solution by $e^{-\beta\epsilon u_0}$ as described
in Ref.~[29]%\citenum{jweek71}
to smooth the first peak around $r = d$.
The solution has better agreement with $\tilde c(m)$
at small wavenumber, but still
predicts a first peak that is lower than simulation
(and lower than the PY solution used in the main text).

  However, it is a good approximation for predicting the correlation
length.  It can be found from Eq.~20 in the main text, using
\begin{align}
1 - \rho c_0 &\equiv \rho Q^{-1}(0) = \frac{\beta}{\rho\kappa} = 1 - 4\pi\rho \int_0^\infty r^2 c(r) dr \label{e:compr} \\
&= \frac{(2\eta + 1)^2}{(1 - \eta)^4} - \tfrac{32}{9} \pi\rho\beta\epsilon \sqrt{2} \label{e:cxn} \\
\rho c_2 &= \frac{4\pi\rho}{6} \int_0^\infty r^4 c(r) dr
= \tfrac{48}{35} \pi\rho\beta\epsilon 2^{5/6} - \frac{d^2}{(1-\eta)^4}\Big(  \\
&\quad (2\eta + 1)^2 (\tfrac{\eta}{20} - \tfrac{3}{16}) + 3(\eta + \tfrac{1}{8})/2
\Big) 
.
\end{align}
The second part in Eq.~\ref{e:cxn} is the zeroth-order perturbation, $\int u_{LR}(r) dr$,
which matches the corresponding term
in the WCA perturbation energy.[29]%\cite{jweek71}

\begin{figure}
\includegraphics[width=0.5\textwidth]{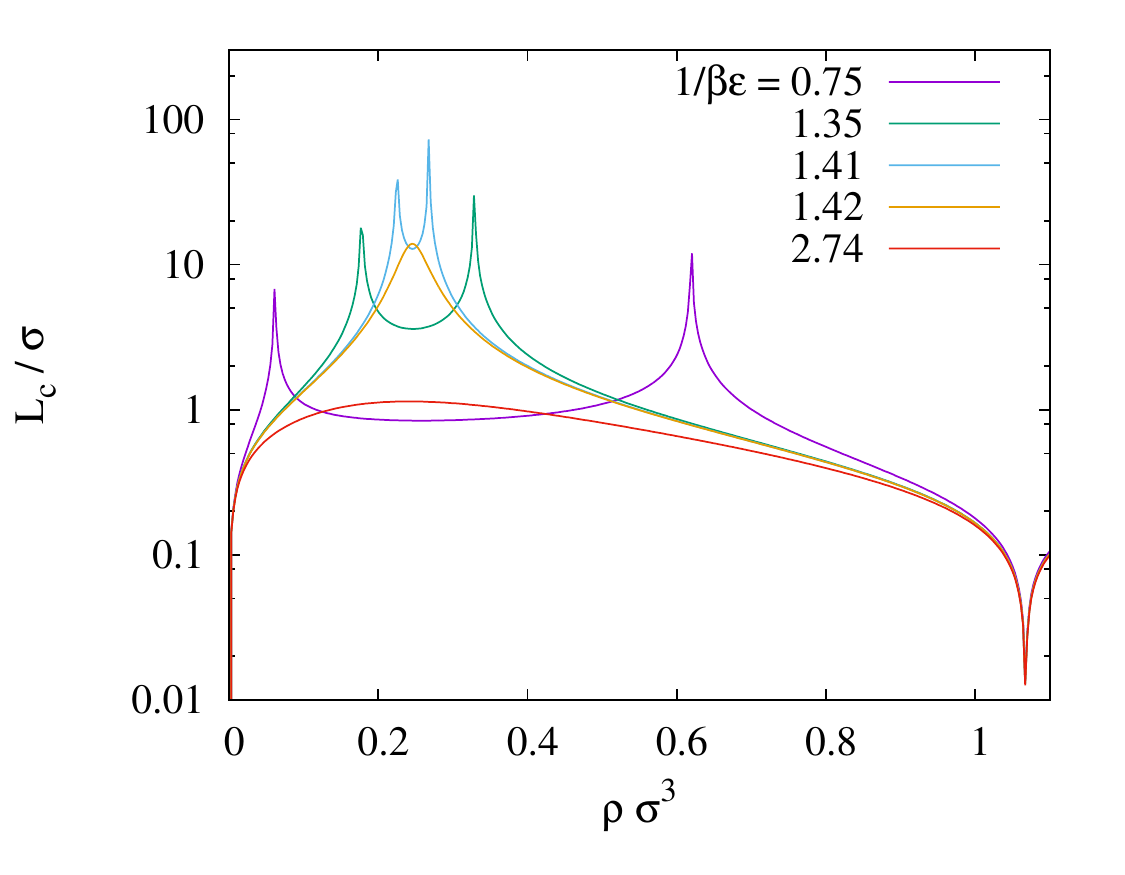}
\caption{Correlation length of the LJ system predicted
from LR perturbation, Eqns.~\ref{e:compr}-\ref{e:cxn}.
The correlation length increases sharply around phase transition
points.  The model predicts a critical temperature
that is too high and a liquid density line that is
too dilute.}\label{f:LJcorr}
\end{figure}

  Fig.~\ref{f:LJcorr} plots Eq.~\ref{e:compr}-\ref{e:cxn}
for the correlation length along a few isotherms.
The numerical value at $\beta^{-1} = 1.5\epsilon$
and $\rho = 0.36\sigma^{-3}$ is $L_c = 2 \sigma$,
which is higher than simulation, but smaller than the PY
approximation.
The correlation length shows the basic features of the phase
diagram, which is remarkable because the hard-sphere
system does not have a liquid-vapor transition.

  Fig.~\ref{f:LJcorr} is based on an approximation
to the WCA model, so its phase behavior is only qualitative.
The qualitative agreement, however, allows us to conclude
that Eq.~\ref{e:mod} is a good approximation to
the small-wavenumber behavior of $\tilde c(m)$
and its inclusion will aid extraction of $\tilde c(m)$
from simulation.

  It also shows that $L_c$ is only few molecular diameters
away from phase transitions.  This short-ranged behavior translates
to a small second derivative of $Q$, which makes estimation of $Q(0)$
possible from the method in the main text
without corrections under ordinary circumstances.

\end{document}